\documentclass[aps,prl,twocolumn,showpacs,superscriptaddress,floatfix]{revtex4-2}  % for review and submission
\usepackage{ucs}
\usepackage{natbib}
\usepackage{graphicx}  % needed for figures
\usepackage{bm}        % for math
\usepackage{amssymb}   % for math
\usepackage{amsmath}
\usepackage{color}
\usepackage{CJK}
\usepackage{times}
\usepackage{verbatim}
\usepackage{mathrsfs}
\usepackage{hyperref}
\usepackage{ulem}
\usepackage{physics}
\hypersetup{colorlinks = True, urlcolor=blue, linkcolor=blue, citecolor=blue}
\newcommand{\br}{\mathbf{r}}

\newcommand{\bg}{\begin{pmatrix}}
\newcommand{\ed}{\end{pmatrix}}
\newcommand{\dg}{\dagger}
\newcommand{\sg}{\sigma}

\usepackage{cancel}
\usepackage{comment}
\begin{document}

\title{Vacancy Spectroscopy of Non-Abelian Kitaev Spin Liquids}
%\input author_list.tex
% D0 authors (remove the first 3 lines
% of this file prior to submission, they
% contain a time stamp for the authorlist)
% (includes institutions and visitors)
\author{Wen-Han Kao}
%\email{kao00018@umn.edu}
\affiliation{School of Physics and Astronomy, University of Minnesota, Minneapolis, MN 55455, USA}

\author{Natalia B. Perkins}
%\email{nperkins@umn.edu}
\affiliation{School of Physics and Astronomy, University of Minnesota, Minneapolis, MN 55455, USA}

\author{G\'abor B. Hal\'asz}
\thanks{This manuscript has been authored in part by UT-Battelle, LLC, under contract DE-AC05-00OR22725 with the US Department of Energy (DOE). The publisher acknowledges the US government license to provide public access under the DOE Public Access Plan (http://energy.gov/downloads/doe-public-access-plan).}
\affiliation{Materials Science and Technology Division, Oak Ridge National Laboratory, Oak Ridge, TN 37831, USA}
\affiliation{Quantum Science Center, Oak Ridge, TN 37831, USA}

\date{\today}
\begin{abstract}
Spin vacancies in the non-Abelian Kitaev spin liquid are known to harbor Majorana zero modes, potentially enabling topological quantum computing at elevated temperatures. Here, we study the spectroscopic signatures of such Majorana zero modes in a scanning tunneling setup where a non-Abelian Kitaev spin liquid with a finite density of spin vacancies forms a tunneling barrier between a tip and a substrate. Our key result is a well-defined peak close to zero bias voltage in the derivative of the tunneling conductance whose voltage and intensity both increase with the density of vacancies. This ``quasi-zero-voltage peak'' is identified as the closest analog of the zero-voltage peak observed in topological superconductors that additionally reflects the fractionalized nature of spin-liquid-based Majorana zero modes. We further highlight a single-fermion Van Hove singularity at a higher voltage that reveals the energy scale of the emergent Majorana fermions in the Kitaev spin liquid. Our proposed signatures are within reach of current experiments on the candidate material $\alpha$-RuCl$_3$.
\end{abstract}
\maketitle

\textit{Introduction.}---The non-Abelian Kitaev spin liquid~\cite{Kitaev2006} is a topologically ordered magnetic insulator that hosts nonlocal fractionalized excitations, including Ising anyons with Majorana zero modes attached to them. Like in topological superconductors~\cite{Alicea2012, Sarma2015}, the non-Abelian particle statistics of such Majorana zero modes may enable intrinsically fault-tolerant topological quantum computation~\cite{Kitaev2003, Nayak2008}. Remarkably though, in promising candidate materials of the non-Abelian Kitaev spin liquid, such as $\alpha$-RuCl$_3$~\cite{Plumb2014, Sandilands2015, Sears2015, Majumder2015, Johnson2015, Sandilands2016, Banerjee2016, Banerjee2017, Do2017} under an in-plane magnetic field~\cite{Kubota2015, Leahy2017, Sears2017, Wolter2017, Baek2017, Banerjee2018, Hentrich2018, Jansa2018, Kasahara2018, Widmann2019, Balz2019, Yamashita2020, Czajka2021, Yokoi2021, Bruin2022, Czajka2023}, the relevant gap is expected to be much larger, thus potentially allowing topological quantum computation at elevated temperatures above $1$ K.

In topological superconductors, Majorana zero modes are bound to localized defects like nanowire end points~\cite{Kitaev2001, Lutchyn2010, Oreg2010} or magnetic vortex cores~\cite{Volovik1999, Read2000, Ivanov2001}, and their standard signature is a zero-voltage peak in the tunneling conductance~\cite{Alicea2012, Sarma2015}, as measured by scanning tunneling microscopy (STM). In the non-Abelian Kitaev spin liquid, the Ising anyons supporting Majorana zero modes can also be bound to pointlike defects such as nonmagnetic spin vacancies~\cite{Willans2010, Willans2011, Kao2021vacancy, Kao2021localization, Vitor2022, comp}, which can be introduced in controlled concentrations into candidate materials like $\alpha$-RuCl$_3$~\cite{Lampen2017, Do2018, Do2020, Imamura2023}. It has also been recently demonstrated that the magnetic excitations of monolayer $\alpha$-RuCl$_3$ can be effectively probed by inelastic electron tunneling spectroscopy where a magnetic insulator forms a tunneling barrier between two metallic electrodes~\cite{Yang2023, Miao2023}. Nevertheless, while related inelastic STM setups with the non-Abelian Kitaev spin liquid acting as a tunneling barrier~\cite{Konig2020} have been theoretically studied to understand signatures of chiral edge modes~\cite{Knolle2020}, bare Ising anyons~\cite{Pereira2020, Udagawa2021, Bauer2023}, and even isolated pairs of spin vacancies~\cite{takahashi2023nonlocal}, it is not yet clear whether the zero-voltage peak as the standard signature of Majorana zero modes has any direct analog in the context of spin liquids.

In this Letter, we answer this question in the affirmative by studying the inelastic STM response for a finite concentration of spin vacancies in the non-Abelian phase of the exactly solvable Kitaev honeycomb model~\cite{Kitaev2006}. Our main result is that the derivative of the tunneling conductance, $\mathrm{d}G / \mathrm{d}V = \mathrm{d}^2 I / \mathrm{d}V^2$, through a spin vacancy in the non-Abelian Kitaev spin liquid contains a well-defined peak close to zero bias voltage $V = 0$ whose characteristic voltage and intensity both increase with the vacancy concentration $n_{\mathrm{v}}$. This ``quasi-zero-voltage peak'' corresponds to the simultaneous excitation of two low-energy fermion modes formed by Majorana zero modes attached to the spin vacancies. While the increase in the peak voltage at a larger vacancy concentration (i.e., smaller average distance between vacancies and, hence, Majorana zero modes) directly carries over from topological superconductors~\cite{Albrecht2016}, the corresponding increase in the peak intensity is a fundamentally new feature reflecting the fractionalized nature of spin-liquid-based Majorana zero modes. We further demonstrate that the controlled introduction of such Majorana zero modes through spin vacancies unveils a single-fermion Van Hove singularity in the STM response of the non-Abelian Kitaev spin liquid, confirming the existence of emergent fermion excitations and revealing their characteristic energy scale.

\begin{figure}
\includegraphics[width=1.0\columnwidth]{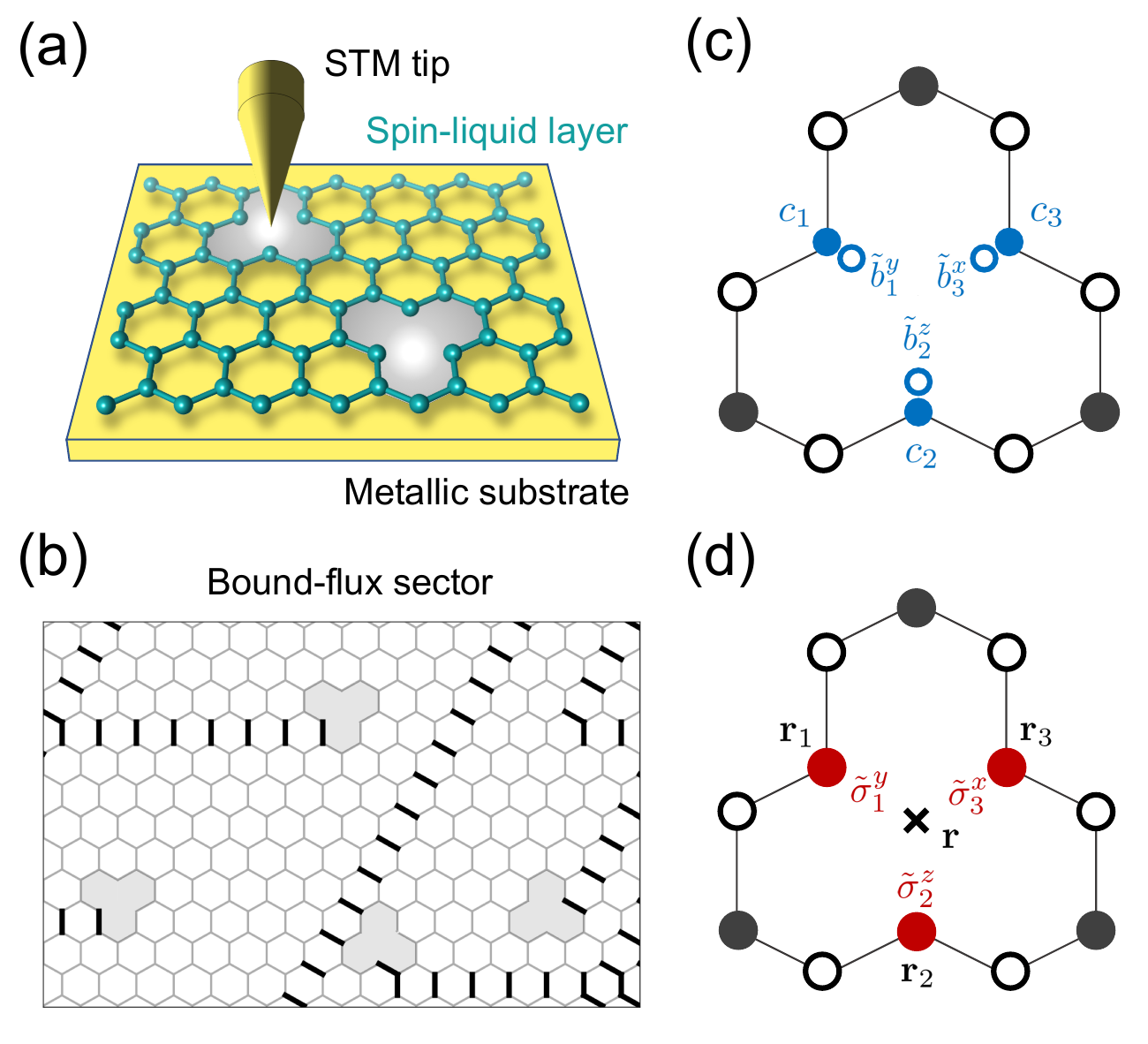}
     \caption{\label{fig:Figure1}(a) Inelastic STM setup for probing local spin dynamics inside the site-diluted Kitaev spin liquid. (b) Bound-flux sector of the site-diluted Kitaev model characterized by flux expectation values $W_p = +1$ (white) and $W_p = -1$ (gray). Thick lines denote strings of bonds $\langle jk \rangle_{\alpha}$ with gauge fields $u_{\langle jk \rangle_{\alpha}} = -1$ that pairwise connect plaquettes $p$ with fluxes $W_p = -1$. (c) Dangling bond fermions $\tilde{b}^{\alpha}_j$ associated with broken bonds around a spin vacancy and the corresponding matter fermions $c_j$ coupled to them by the third term of Eq.~(\ref{eq:H-3}). (d) Dangling spin components $\tilde{\sg}^{\alpha}_j = i \tilde{b}^{\alpha}_j c_j$ that fractionalize into dangling bond fermions $\tilde{b}^{\alpha}_j$ and matter fermions $c_j$.}
\end{figure}

\textit{Scanning tunneling formalism.}---We consider the setup in Fig.~\ref{fig:Figure1}(a) where a non-Abelian Kitaev spin liquid with site dilution (i.e., a finite concentration of spin vacancies) forms a tunneling barrier between a metallic STM tip and a metallic substrate~\cite{Konig2020, Knolle2020, Pereira2020, Udagawa2021, Bauer2023, takahashi2023nonlocal}. The Hamiltonian of the system is given by $\mathcal{H} = \mathcal{H}_{\mathrm{tip}} + \mathcal{H}_{\mathrm{sub}} + \mathcal{H}_{\mathrm{Kitaev}} + \mathcal{H}_{\mathrm{tunnel}}$, where
\begin{align}
\mathcal{H}_{\mathrm{tip}} = \sum_{\mathbf{p}\sg} \hat{\varepsilon}^{\phantom{\dg}}_{\mathbf{p}} \hat{c}^{\dg}_{\mathbf{p}\sg} \hat{c}^{\phantom{\dg}}_{\mathbf{p}\sg}, \quad \mathcal{H}_{\mathrm{sub}} = \sum_{\mathbf{k}\sg} \check{\varepsilon}_{\mathbf{k}} \check{c}^{\dg}_{\mathbf{k}\sg} \check{c}^{\phantom{\dg}}_{\mathbf{k}\sg}, \label{eq:H-1}
\end{align}
are free-electron Hamiltonians for the tip and the substrate, respectively, $\mathcal{H}_{\mathrm{Kitaev}}$ is the spin Hamiltonian of the site-diluted Kitaev spin liquid (to be specified later), while
\begin{align}
\mathcal{H}_{\mathrm{tunnel}} = \sum_{\mathbf{k}\mathbf{p}\mu\nu} T^{\phantom{\dg}}_{\mu\nu} (\mathbf{r}) \, \check{c}^{\dg}_{\mathbf{k}\mu} \hat{c}^{\phantom{\dg}}_{\mathbf{p}\nu} + \mathrm{H.c.} \label{eq:H-2}
\end{align}
describes the tunneling of electrons between the tip and the substrate through the spin liquid as a function of the tip position $\mathbf{r}$. The tunneling matrix element can be written as
\begin{align}
T^{\phantom{\dg}}_{\mu\nu} (\mathbf{r}) = \sum_{j} \left[ T^{0}_{\mu\nu} (\mathbf{r} - \mathbf{r}_j) + \vec{\sg}_j \cdot \vec{T}_{\mu\nu} (\mathbf{r} - \mathbf{r}_j) \right], \label{eq:T-1}
\end{align}
where $\vec{\sg}_j$ is the spin at location $\mathbf{r}_j$ inside the spin liquid, while $T^{0}_{\mu\nu}$ and $\vec{T}_{\mu\nu} = [T^{x}_{\mu\nu}, T^{y}_{\mu\nu}, T^{z}_{\mu\nu}]$ are generally given by~\cite{Bauer2023}
\begin{align}
& T^{0}_{\mu\nu} (\mathbf{r} - \mathbf{r}_j) = T (\mathbf{r} - \mathbf{r}_j) \, \eta_0 \delta_{\mu\nu}, \label{eq:T-2} \\
& T^{\alpha}_{\mu\nu} (\mathbf{r} - \mathbf{r}_j) = T (\mathbf{r} - \mathbf{r}_j) \left[ \eta_1 \tau^{\alpha}_{\mu\nu} + \eta_2 \left( \tau^{x}_{\mu\nu} + \tau^{y}_{\mu\nu} + \tau^{z}_{\mu\nu} \right) \right] \nonumber
\end{align}
with $\alpha = x,y,z$ and the Pauli matrices $\tau^{x,y,z}$. The real $O(1)$ coefficients $\eta_{0,1,2}$ depend on microscopic details~\cite{Bauer2023}, and the overall tunneling amplitude through site $j$ of the spin liquid is $T (\mathbf{r} - \mathbf{r}_j) = t_{\mathrm{tip}} (\mathbf{r} - \mathbf{r}_j) \, t_{\mathrm{sub}} / U$, where $U$ is the charge gap of the spin liquid, and $t_{\mathrm{tip}}$ ($t_{\mathrm{sub}}$) is the hopping amplitude between site $j$ and the tip (substrate). Since the tip is atomically sharp, $t_{\mathrm{tip}}$ decays exponentially with the distance $|\mathbf{r} - \mathbf{r}_j|$ between the tip and the given site $j$. Without affecting our main results, we set $\eta_2 = 0$ in the rest of this work~\cite{Knolle2020}.

In the presence of a bias voltage $V$ between the tip and the substrate, the resulting tunneling current $I$ can be computed via Fermi's golden rule. Assuming constant densities of states $D_{\mathrm{tip}}$ and $D_{\mathrm{sub}}$ for both the tip and the substrate, the differential conductance $G(V) = \mathrm{d}I / \mathrm{d}V$ at zero temperature can be written as (see the SM~\cite{supp} for detailed derivation)
\begin{align}
G(V) = \frac{2\pi e^2} {\hbar} D_{\mathrm{tip}} D_{\mathrm{sub}} \left[ K_{\mathrm{elastic}} + \int_{0}^{eV} d\omega \, K_{\mathrm{inelastic}} (\omega) \right] \label{eq:G}
\end{align}
in terms of an elastic and an inelastic component. The kernel of the inelastic component takes the form~\cite{supp}
\begin{align}
K_{\mathrm{inelastic}} (\omega)
= 2\eta_1^2 \sum_{jk}\sum_{\alpha}  T(\br-\br_j) \, T(\br-\br_k) \, S^{\alpha\alpha}_{jk} (\omega), \label{eq:S-1}
\end{align}
where the sum in $\alpha$ runs over $x,y,z$, and $S^{\alpha\beta}_{jk} (\omega)$ is the connected dynamical spin correlation function:
\begin{align}
S^{\alpha\beta}_{jk} (\omega) = \frac{1} {2\pi} \int_{-\infty}^{+\infty} dt \, e^{i \omega t} \left[ \langle \sg^{\alpha}_j (t) \sg^{\beta}_k (0) \rangle - \langle \sg^{\alpha}_j \rangle \langle \sg^{\beta}_k \rangle \right]. \label{eq:S-2}
\end{align}
Therefore, the derivative of the tunneling conductance at finite voltage, $\mathrm{d}G(V) / \mathrm{d}V = \mathrm{d}^2 I / \mathrm{d}V^2 \propto K_{\mathrm{inelastic}} (eV)$, is sensitive to the local spin dynamics of the site-diluted Kitaev spin liquid at a given frequency $\omega = eV$.

\textit{Site-diluted Kitaev spin liquid.}---We introduce a finite concentration of nonmagnetic spin vacancies into a non-Abelian Kitaev spin liquid by removing the spins $\vec{\sg}_j$ from a set of randomly selected vacancy sites $j \in \mathbb{V}$. For the remaining sites, we take the simplest possible exactly solvable Hamiltonian,
\begin{align}
\mathcal{H}_{\mathrm{Kitaev}} = -J \sum_{\langle jk \rangle_{\alpha}} \sg^{\alpha}_{j} \sg^{\alpha}_{k} - \kappa \sum_{\langle jkl \rangle_{\alpha \beta}} \sg^{\alpha}_{j} \sg^{\gamma}_{k} \sg^{\beta}_{l} - h \sum_{j \in \mathbb{D}_{\alpha}} \sg^{\alpha}_{j}, \label{eq:H-3}
\end{align}
where $\mathbb{D}_{\alpha}$ is the set of sites connected to a vacancy site by an $\alpha$ bond $\langle jk \rangle_{\alpha}$, and $\langle jkl \rangle_{\alpha \beta}$ is the path consisting of the bonds $\langle jk \rangle_{\alpha}$ and $\langle kl \rangle_{\beta}$ with $(\alpha\beta\gamma)$ being a general permutation of $(xyz)$. The first term represents the pure Kitaev honeycomb model~\cite{Kitaev2006}, and the remaining terms describe a magnetic field that breaks time-reversal symmetry. The second term perturbatively incorporates a small magnetic field in the bulk of the spin liquid while preserving the exact solution of the pure Kitaev model~\cite{Kitaev2006}. At each site $j \in \mathbb{D}_{\alpha}$, however, the $\alpha$ component of the magnetic (i.e., Zeeman) field can also be directly included, as seen in the third term of Eq.~(\ref{eq:H-3}), and the resulting Hamiltonian is still exactly solvable~\cite{Kao2021vacancy, takahashi2023nonlocal, comp}.

This property can be understood by employing the standard four-Majorana representation of the spins, $\sg^{\alpha}_j = i b^{\alpha}_j c_j$, with the local parity constraint $b^x_j b^y_j b^z_j c_j = 1$~\cite{Kitaev2006}. In terms of the Majorana fermions $b^{\alpha}_j$ and $c_j$, the first two terms of Eq.~(\ref{eq:H-3}) read $iJ \sum_{\langle jk \rangle_{\alpha}} u_{\langle jk \rangle_{\alpha}} c_j c_k$ and $i\kappa \sum_{\langle jkl \rangle_{\alpha \beta}} u_{\langle jk \rangle_{\alpha}} u_{\langle kl \rangle_{\beta}} c_j c_l$, where $u_{\langle jk \rangle_{\alpha}} = i b^{\alpha}_j b^{\alpha}_k = \pm 1$ are conserved quantities. Therefore, one obtains a quadratic hopping problem for the matter fermions $c_j$ that move in the presence of static $\mathbb{Z}_2$ gauge fields $u_{\langle jk \rangle_{\alpha}}$. Including a Zeeman field $\propto i b^{\alpha}_j c_j$ at a generic site $j$, the corresponding gauge field $u_{\langle jk \rangle_{\alpha}}$ is then no longer a conserved quantity because the bond fermion $b^{\alpha}_j$ anticommutes with it. For a site $j \in \mathbb{D}_{\alpha}$, however, there is no gauge field $u_{\langle jk \rangle_{\alpha}}$ because the site $k \in \mathbb{V}$ is removed. Instead, $\tilde{b}^{\alpha}_j \equiv b^{\alpha}_j$ is a ``dangling'' (unpaired) bond fermion [see Fig.~\ref{fig:Figure1}(c)] that does not appear in the first two terms of Eq.~(\ref{eq:H-3}) at all. Therefore, the Zeeman fields $\propto i \tilde{b}^{\alpha}_j c_j$ in the third term of Eq.~(\ref{eq:H-3}) simply incorporate the dangling bond fermions $\tilde{b}^{\alpha}_j$ into the quadratic hopping problem for the matter fermions $c_j$. Note that, in the rest of this work, we use a tilde to distinguish the dangling bond fermions $\tilde{b}^{\alpha}_j$ and the dangling spin components $\tilde{\sg}^{\alpha}_j = i \tilde{b}^{\alpha}_j c_j$ associated with them [see Fig.~\ref{fig:Figure1}(d)].

The elementary excitations of the Hamiltonian $\mathcal{H}_{\mathrm{Kitaev}}$ in Eq.~(\ref{eq:H-3}) are the matter fermions $c_j$ (which include the dangling bond fermions $\tilde{b}^{\alpha}_j$) and the $\mathbb{Z}_2$ gauge fluxes around the plaquettes $p$ of the lattice: $W_p = \prod_{\langle jk \rangle_{\alpha} \in p} u_{\langle jk \rangle_{\alpha}} = \pm 1$. For the clean model without any vacancies, the ground state is in the zero-flux sector~\cite{Kitaev2006, Lieb1994} where $W_p = +1$ for all $p$. In the site-diluted model, however, if the vacancy concentration and the magnetic field are sufficiently small, the ground state belongs to the bound-flux sector~\cite{Willans2010, Willans2011, Kao2021vacancy, Kao2021localization, Vitor2022, comp} where $W_p = +1$ for the elementary hexagonal plaquettes but $W_p = -1$ for the enlarged vacancy plaquettes formed by three hexagonal plaquettes around each vacancy site [see Fig.~\ref{fig:Figure1}(b)]. In turn, each plaquette with $W_p = -1$ corresponds to a non-Abelian Ising anyon~\cite{Kitaev2006} that supports a localized Majorana zero mode in the hopping problem of the matter fermions. Therefore, the introduction of vacancies can stabilize Majorana zero modes in the ground state of the non-Abelian Kitaev spin liquid.

\textit{Local spin dynamics.}---The tunneling conductance through the site-diluted Kitaev spin liquid depends on the dynamical spin correlation function $S^{\alpha\beta}_{jk} (\omega)$ in Eq.~(\ref{eq:S-2}). This quantity can be written in the Lehmann representation as
\begin{align}
S^{\alpha\beta}_{jk} (\omega) = \sum_{\Phi} \langle \Omega | \sg^{\alpha}_j | \Phi \rangle \langle \Phi | \sg^{\beta}_k | \Omega \rangle \, \delta (\omega + E_{\Omega} - E_{\Phi}), \label{eq:S-3}
\end{align}
where $| \Omega \rangle$ is the ground state of $\mathcal{H}_{\mathrm{Kitaev}}$ in Eq.~(\ref{eq:H-3}) with energy $E_{\Omega}$, while $| \Phi \rangle$ are excited states with energies $E_{\Phi}$. For generic sites $j$ and $k$, the relevant excited states contain a pair of flux excitations, and $S^{\alpha\beta}_{jk} (\omega)$ can be computed following Refs.~\cite{Baskaran2007, Knolle2014, Knolle2015, Zschocke2015}. For $j \in \mathbb{D}_{\alpha}$ and $k \in \mathbb{D}_{\beta}$, however, the relevant excited states $| \Phi \rangle$ only contain two matter-fermion excitations, and the matrix elements $\langle \Phi | \tilde{\sg}^{\alpha}_j | \Omega \rangle$ and $\langle \Phi | \tilde{\sg}^{\beta}_k | \Omega \rangle$ of the dangling spin components $\tilde{\sg}^{\alpha}_j = i \tilde{b}^{\alpha}_j c_j$ and $\tilde{\sg}^{\beta}_k = i \tilde{b}^{\beta}_k c_k$ become standard four-fermion expectation values. The details of computing $S^{\alpha\beta}_{jk} (\omega)$ are described in the SM~\cite{supp}.

\begin{figure*}
\includegraphics[width=1.0\textwidth]{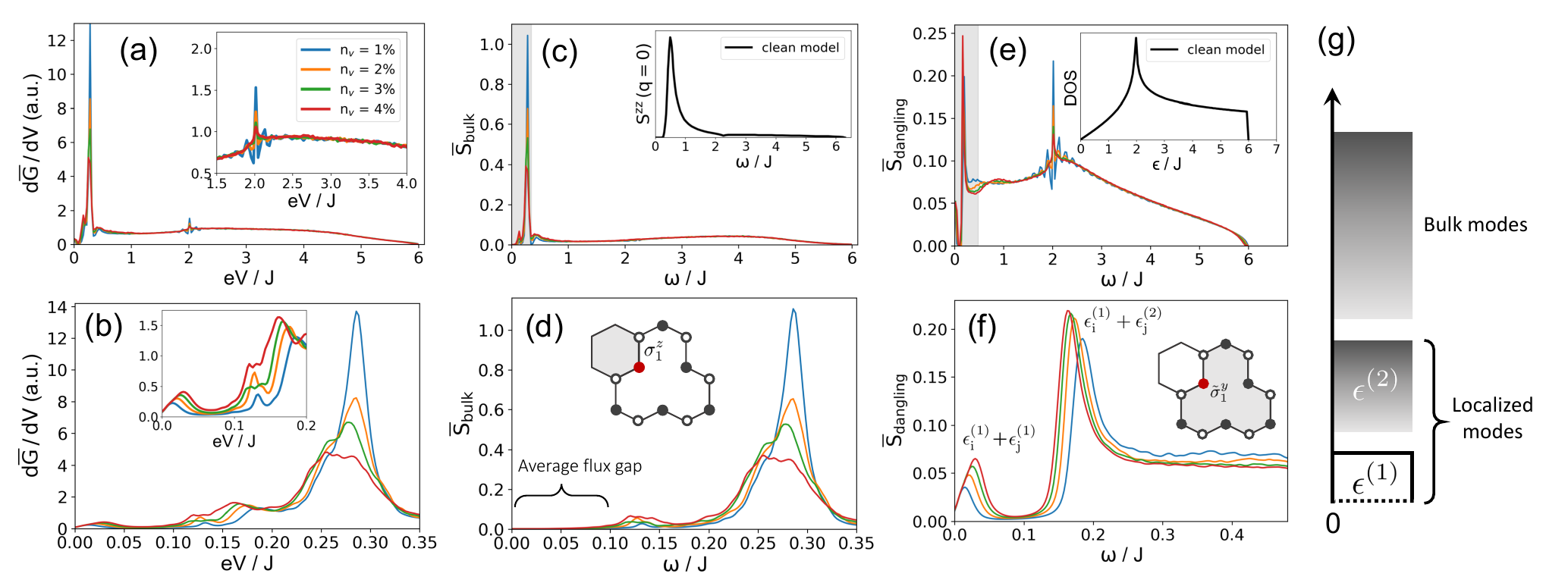}
     \caption{\label{fig:Figure2} (a,b) Disorder-averaged derivative of the tunneling conductance against the bias voltage for vacancy concentrations $1\% \leq n_{\mathrm{v}} \leq 4\%$ in the entire voltage range (a) and at low voltages (b). The two insets emphasize specific voltage ranges. (c-f) Disorder-averaged bulk (c,d) and dangling (e,f) spin correlation functions against the energy transfer for vacancy concentrations $1\% \leq n_{\mathrm{v}} \leq 4\%$ in the entire energy range (c,e) and at low energies (d,f). The insets of (c) and (e) show the zero-momentum spin correlation function and the matter-fermion density of states in the clean model, respectively, while the insets of (d) and (f) depict the relevant spin components and the flux sectors of the corresponding intermediate states $| \Phi \rangle$ in Eq.~(\ref{eq:S-3}) with flux expectation values $W_p = +1$ ($W_p = -1$) marked by white (gray). Note that the insets of (d) and (f) have two and zero flux excitations with respect to the bound-flux sector [see Fig.~\ref{fig:Figure1}(b)], respectively. Panels (a-f) are all based on Eq.~(\ref{eq:H-3}) with $\kappa = 0.02J$ and $h = 0.05J$, involve disorder averaging over $500$ realizations in a $40 \times 40$ system, and use the same color scheme for $n_{\mathrm{v}}$. (g) Schematic energy spectrum of the quadratic fermion problem involving the matter fermions $c_j$ and the dangling bond fermions $\tilde{b}^{\alpha}_j$ with dominant $c_j$ ($\tilde{b}^{\alpha}_j$) character marked by shading (no shading). The peak energies in (f) are given in terms of the fermion energies $\epsilon^{(1,2)}$.}
\end{figure*}

\textit{Results and discussion.}---Figure~\ref{fig:Figure2} shows the derivative of the tunneling conductance, $\mathrm{d}G / \mathrm{d}V \propto K_{\mathrm{inelastic}} (eV)$, as obtained from Eqs.~(\ref{eq:S-1}) and (\ref{eq:S-2}), against the inelastic energy transfer $\omega = eV$ into the site-diluted Kitaev spin liquid. Without affecting our main results, we assume that the STM tip is directly above a vacancy site [see Figs.~\ref{fig:Figure1}(a,d)] and that, due to the atomically sharp nature of the tip, the only contributions to the tunneling conductance are from the three neighboring sites $j = 1,2,3$ [see Fig.~\ref{fig:Figure1}(d)] with identical tunneling amplitudes $T (\br - \br_j)$. To obtain a typical result at a finite vacancy concentration $1\% \leq n_{\mathrm{v}} \leq 4\%$, we also perform a disorder average over all randomly positioned vacancies in a $40 \times 40$ system and over $500$ disorder realizations with different randomly selected vacancy positions.

The disorder-averaged derivative of the tunneling conductance shown in Figs.~\ref{fig:Figure2}(a,b) can be interpreted as the sum of two components, ``bulk'' and ``dangling'', that are plotted in Figs.~\ref{fig:Figure2}(c,d) and \ref{fig:Figure2}(e,f), respectively. To understand these components, we first recognize from Eq.~(\ref{eq:S-3}) that $S^{\alpha\alpha}_{jk} (\omega)$ is only nonzero for $j=k$ among the sites $j,k = 1,2,3$ surrounding each vacancy~\cite{comp}. Thus, we can use Eq.~(\ref{eq:S-1}) to write the disorder-averaged derivative of the tunneling conductance as
\begin{align}
\frac{\mathrm{d}\overline{G}(V)}{\mathrm{d}V} &\propto \sum_{\alpha} \sum_{j=1}^3 \overline{S^{\alpha\alpha}_{jj}} (eV) \label{eq:GV-1} \\
&= 6 \, \overline{S}_{{\mathrm{bulk}}} (eV) + 3 \, \overline{S}_{\mathrm{dangling}} (eV), \nonumber
\end{align}
where the two symmetry-inequivalent disorder-averaged spin correlation functions,
\begin{align}
\begin{split}
& \overline{S}_{\mathrm{bulk}} (\omega) = \overline{S^{\alpha\alpha}_{jj}} (\omega) \qquad \,\, (j \notin \mathbb{D_{\alpha}}), \label{eq:GV-2} \\
& \overline{S}_{\mathrm{dangling}} (\omega) = \overline{S^{\alpha\alpha}_{jj}} (\omega) \quad (j \in \mathbb{D_{\alpha}}),
\end{split}
\end{align}
involve non-dangling and dangling spin components, respectively. Note that the overline means disorder average.

The distinct structures of the bulk and dangling components can be understood by referring to the Lehmann representation in Eq.~(\ref{eq:S-3}). In the bulk spin correlation function $\overline{S}_{\mathrm{bulk}} (\omega)$,  each of the relevant intermediate states $| \Phi \rangle$  contains a pair of flux excitations at neighboring plaquettes [see inset of Fig.~\ref{fig:Figure2}(d)], like in the clean model~\cite{Baskaran2007, Knolle2014, Knolle2015, Zschocke2015}. Hence, $\overline{S}_{\mathrm{bulk}} (\omega)$ largely follows the spin correlation function of the clean model [see inset of Fig.~\ref{fig:Figure2}(c)], exhibiting a finite flux gap, a strong peak above the flux gap, and a diffuse continuum at higher energies~\footnote{Note that the flux gap is smaller than in the clean model because a flux excitation at the vacancy plaquette costs less energy.}. In the dangling spin correlation function $\overline{S}_{\mathrm{dangling}} (\omega)$, however, the intermediate states $| \Phi \rangle$ do not contain any flux excitations [see inset of Fig.~\ref{fig:Figure2}(f)]. Thus, $\overline{S}_{\mathrm{dangling}} (\omega)$ is only sensitive to correlations involving matter fermions $c_j$ and dangling bond fermions $\tilde{b}^{\alpha}_j$ in the bound-flux sector~\cite{comp}.

Specifically, the structure of $\overline{S}_{\mathrm{dangling}} (\omega)$ [see Figs.~\ref{fig:Figure2}(e,f)] reflects the spectrum of the quadratic fermion problem for $c_j$ and $\tilde{b}^{\alpha}_j$, which is schematically shown in Fig.~\ref{fig:Figure2}(g). Below the delocalized (``bulk'') fermion modes inherited from the clean model, there are two bands of localized modes induced by the vacancies with an energy gap in between them. While the bulk modes are of predominantly $c_j$ character, the localized modes have significant contributions from both $c_j$ and $\tilde{b}^{\alpha}_j$ with the upper (lower) modes dominated by $c_j$ ($\tilde{b}^{\alpha}_j$)~\cite{comp}. Thus, recalling $\tilde{\sg}^{\alpha}_j = i \tilde{b}^{\alpha}_j c_j$, the intermediate states $| \Phi \rangle$ with the largest matrix elements $\langle \Phi | \tilde{\sg}^{\alpha}_j | \Omega \rangle$ in Eq.~(\ref{eq:S-3}) contain one fermion from a lower localized mode and one fermion from either a higher localized mode or a bulk mode. The strong peak at $\omega \approx 0.2J$ [see Fig.~\ref{fig:Figure2}(f)] can then be attributed to two localized fermion excitations (one from the upper and lower bands each), while the continuum at $0.2J \lesssim \omega \lesssim 6J$ [see Fig.~\ref{fig:Figure2}(e)] corresponds to one bulk excitation as well as one localized excitation from the lower band. Since such a localized excitation costs almost no energy (less than $0.1J$) and the bulk excitation is inherited from the clean model, the intensity of the continuum largely follows the \textit{single-fermion} density of states in the clean model [see inset of Fig.~\ref{fig:Figure2}(e)] with a pronounced Van Hove singularity at $\omega \approx 2J$~\footnote{Note that there are discrepancies due to matrix-element effects and overlap with the peaks at lower energies.}. This result is remarkable because the matter fermions are fractionalized excitations and no local probe can directly couple to a single one of them.

Nevertheless, the most intriguing feature of $\overline{S}_{\mathrm{dangling}} (\omega)$ is the low-energy peak at $\omega \lesssim 0.05J$ whose energy and intensity both increase as the vacancy concentration is tuned from $1\%$ to $4\%$ [see Fig.~\ref{fig:Figure2}(f)]. This peak corresponds to two localized fermion excitations from the lower band and is relatively weak because both excitations are of predominantly $\tilde{b}^{\alpha}_j$ character and a product of two $\tilde{b}^{\alpha}_j$ fermions has little overlap with $\tilde{\sg}^{\alpha}_j = i \tilde{b}^{\alpha}_j c_j$. Furthermore, due to the Pauli exclusion principle, the intensity of this peak is only nonzero if two distinct localized fermion modes from the lower band have finite support on the same site $j$, which in turn requires hybridization between dangling bond fermions $\tilde{b}^{\alpha}_j$ at different sites $j$. While each vacancy has three $\tilde{b}^{\alpha}_j$ fermions connected to it, the hybridization between them is weak due to the large distance around the vacancy [see Fig.~\ref{fig:Figure1}(c)]. However, as the vacancy concentration increases, the $\tilde{b}^{\alpha}_j$ fermions connected to different vacancies can also start hybridizing. Therefore, by recognizing that  $\tilde{b}^{\alpha}_j$ fermions are at exactly zero energy without any hybridization between them, we can immediately understand why the energy and the intensity of the low-energy peak both increase with the vacancy concentration $n_{\mathrm{v}}$.

From a more universal perspective, the dangling Majorana fermions $\tilde{b}^{\alpha}_j$ can be interpreted as Majorana zero modes. The existence of a low-energy peak and the $n_{\mathrm{v}}$ dependence of its energy then reveal the presence of Majorana zero modes~\cite{Albrecht2016}, while the $n_{\mathrm{v}}$ dependence of the peak intensity reflects that Majorana zero modes in spin liquids are fractionalized particles such that they can only couple to a local probe if they are close to each other. Moreover, the direct observability of the single-fermion density of states is enabled by the presence of localized Majorana zero modes costing zero energy. This universal perspective also makes it clear that our results apply to site-diluted Kitaev spin liquids beyond the exactly solvable model in Eq.~(\ref{eq:H-3}) as an Ising anyon in a non-Abelian Kitaev spin liquid necessarily has a Majorana zero mode bound to it~\footnote{The three Majorana zero modes in the exactly solvable model hybridize in the presence of generic perturbations, thus forming a finite-energy complex-fermion mode as well as a single residual Majorana zero mode.}. We finally point out that, while $\mathrm{d}\overline{G} / \mathrm{d}V$ in Figs.~\ref{fig:Figure2}(a,b) is largely dominated by $\overline{S}_{\mathrm{bulk}} (\omega)$ in Figs.~\ref{fig:Figure2}(c,d), the key features in $\overline{S}_{\mathrm{dangling}} (\omega)$, including the single-fermion Van Hove singularity and the $n_{\mathrm{v}}$-dependent low-voltage peak, are clearly visible in $\mathrm{d}\overline{G} / \mathrm{d}V$ [see insets of Figs.~\ref{fig:Figure2}(a,b)].

\textit{Summary.}---We considered the inelastic STM response of a non-Abelian Kitaev spin liquid containing a finite concentration of vacancies that forms a tunneling barrier in between a STM tip and a metallic substrate. We identified a well-defined peak close to zero bias voltage ($eV \ll J$) in the derivative of the tunneling conductance, $\mathrm{d}G / \mathrm{d}V$, whose voltage and intensity both increase with the vacancy concentration, and interpreted this ``quasi-zero-voltage peak'' as a universal signature of spin-liquid-based Majorana zero modes. Further, we found a single-fermion Van Hove singularity at $eV \approx 2J$ that confirms the presence of emergent Majorana fermions and reveals the magnitude of the Kitaev interaction strength $J$~\footnote{In this work, we use the ``$\sigma \cdot \sigma$'' convention for writing the spin Hamiltonian. Thus, the values of $J$ for $\alpha$-RuCl$_3$ reported in the literature (typically using the ``$S \cdot S$'' convention) correspond to the range between $1$ meV and $5$ meV in our convention.}.

\textit{Acknowledgments}.---We thank Jason Alicea, Patrick Lee, and Alan Tennant for enlightening discussions, as well as An-Ping Li for helpful comments on the manuscript. G.~B.~H. was supported by the U.S. Department of Energy, Office of Science, National Quantum Information Science Research Centers, Quantum Science Center. W.-H. Kao and N. B. Perkins acknowledge the support from NSF DMR-2310318 and  the support of the Minnesota Supercomputing Institute (MSI) at the University of Minnesota.

%\onecolumngrid

%\bibliographystyle{apsrev4-1}
\bibliography{STM_refs.bib}
\end{document}

\title{Supplemental Material for "Vacancy Spectroscopy of Non-Abelian Kitaev Spin Liquids"}
%\input author_list.tex
% D0 authors (remove the first 3 lines
% of this file prior to submission, they
% contain a time stamp for the authorlist)
% (includes institutions and visitors)
\author{Wen-Han Kao}
%\email{kao00018@umn.edu}
\affiliation{School of Physics and Astronomy, University of Minnesota, Minneapolis, MN 55455, USA}

\author{Natalia B. Perkins}
%\email{nperkins@umn.edu}
\affiliation{School of Physics and Astronomy, University of Minnesota, Minneapolis, MN 55455, USA}

\author{G\'abor B. Hal\'asz}
\affiliation{Materials Science and Technology Division, Oak Ridge National Laboratory, Oak Ridge, TN 37831, USA}
\affiliation{Quantum Science Center, Oak Ridge, TN 37831, USA}

\date{\today}
%\begin{abstract}

%\end{abstract}
\pacs{}
\maketitle
In this supplemental material, we derive the tunneling current and the differential tunneling conductance of the STM setup in the main text. The inelastic component of the tunneling conductance is found to be proportional to the dynamical spin correlation function of the site-diluted Kitaev honeycomb model. This spin correlation function is also discussed here.

\setcounter{figure}{0}
\setcounter{equation}{0}
\renewcommand{\theequation}{S\arabic{equation}}
\renewcommand{\thefigure}{S\arabic{figure}}

\section{Theory of the tunneling current}\label{sec:tunneling_current}

We compute the tunneling current $I$ from the tip to the substrate in the framework of Fermi's golden rule by treating $\mathcal{H}_{\mathrm{tunnel}}$ as a perturbation on top of $\mathcal{H}_{\mathrm{tip}} + \mathcal{H}_{\mathrm{sub}} + \mathcal{H}_{\mathrm{Kitaev}}$ [see Eqs.~(1), (2), and (8) in the main text]. At zero temperature, the system consisting of the tip, substrate, and spin liquid is initially in its ground state,
\begin{align}
\ket{\Psi} = \ket{\Theta_{\mathrm{tip}}}\otimes \ket{\Theta_{\mathrm{sub}}}\otimes \ket{\Omega},
\end{align}
where $\ket{\Theta_{\mathrm{tip}}}$ and $\ket{\Theta_{\mathrm{sub}}}$ are filled Fermi seas for the tip and the substrate with electrons $(c^{\mathrm{tip}}_{\mathbf{p}\sg})^{\dg} \equiv \hat{c}^{\dg}_{\mathbf{p}\sg}$ and $(c^{\mathrm{sub}}_{\mathbf{k}\sg})^{\dg} \equiv \check{c}^{\dg}_{\mathbf{k}\sg}$ up to the Fermi levels $\varepsilon^{\mathrm{tip}}_{\mathbf{p}} \equiv \hat{\varepsilon}_{\mathbf{p}} = 0$ and $\varepsilon^{\mathrm{sub}}_{\mathbf{k}} \equiv \check{\varepsilon}_{\mathbf{k}} = 0$, respectively, while $\ket{\Omega}$ is the ground state of $\mathcal{H}_{\mathrm{Kitaev}}$ in Eq.~(8) of the main text. After the tunneling process, a general final state can be written as
\begin{align}
\ket{\Psi'} = \ket{\Theta_{\mathrm{tip}}'}\otimes \ket{\Theta_{\mathrm{sub}}'}\otimes \ket{\Xi},
\end{align}
where $\ket{\Theta_{\mathrm{tip}}'}$ and $\ket{\Theta_{\mathrm{sub}}'}$ contain one less and one more electron with respect to $\ket{\Theta_{\mathrm{tip}}}$ and $\ket{\Theta_{\mathrm{sub}}}$, respectively, while $\ket{\Xi}$ is a general eigenstate of $\mathcal{H}_{\mathrm{Kitaev}}$ with energy $E_{\Xi}$ that may be the ground state $\ket{\Omega}$ with energy $E_{\Omega}$ or an excited state $\ket{\Phi}$ with energy $E_{\Phi}$. In the presence of a bias voltage $V$, the tunneling current according to Fermi's golden rule is given by
\begin{align}
I &= \frac{2\pi e}{\hbar} \sum_{\bk\bp\mu\nu} \sum_{\Psi'} \left| \bra{\Psi'} T_{\mu\nu}(\br) (c^{\mathrm{sub}}_{\bk\mu})^{\dg} c^{\mathrm{tip}}_{\bp\nu}  \ket{\Psi} \right|^2 \, \dt(E_{\Xi} - E_{\Omega} + \varepsilon^{\mathrm{sub}}_{\mathbf{k}} - \varepsilon^{\mathrm{tip}}_{\mathbf{p}} - eV) \\
&= \frac{2\pi e}{\hbar} \sum_{\bk\bp\mu\nu} \sum_{\Theta_{\mathrm{tip}}'} \sum_{\Theta_{\mathrm{sub}}'} \sum_{\Xi} \big| \bra{\Theta_{\mathrm{tip}}'} c^{\mathrm{tip}}_{\bp\nu} \ket{\Theta_{\mathrm{tip}}} \big|^2 \, \big| \bra{\Theta_{\mathrm{sub}}'} (c^{\mathrm{sub}}_{\bk\mu})^{\dg} \ket{\Theta_{\mathrm{sub}}} \big|^2 \, \big| \bra{\Xi} T_{\mu\nu}(\br) \ket{\Omega} \big|^2 \, \dt(E_{\Xi} - E_{\Omega} + \varepsilon^{\mathrm{sub}}_{\mathbf{k}} - \varepsilon^{\mathrm{tip}}_{\mathbf{p}} - eV). \nonumber
\end{align}
Taking the continuum limit with $\varepsilon^{\mathrm{tip}}_{\mathbf{p}} \to \varepsilon_{\mathrm{tip}}$ and $\varepsilon^{\mathrm{sub}}_{\mathbf{k}} \to \varepsilon_{\mathrm{sub}}$, this tunneling current then becomes
\begin{align}
I &= \frac{2\pi e}{\hbar} D_{\mathrm{tip}} D_{\mathrm{sub}} \sum_{\mu\nu}\sum_{\Xi} \int d\varepsilon_{\mathrm{tip}} \int d\varepsilon_{\mathrm{sub}} \, \theta (-\varepsilon_{\mathrm{tip}}) \, \theta (\varepsilon_{\mathrm{sub}}) \, \big| \bra{\Xi} T_{\mu\nu}(\br) \ket{\Omega} \big|^2 \, \dt(E_{\Xi} - E_{\Omega} + \varepsilon_{\mathrm{sub}} - \varepsilon_{\mathrm{tip}} - eV) \nonumber \\
&= \frac{2\pi e}{\hbar} D_{\mathrm{tip}} D_{\mathrm{sub}} \sum_{\mu\nu}\sum_{\Xi} \int d\varepsilon_{\mathrm{sub}} \, \theta (\varepsilon_{\mathrm{sub}}) \, \theta (eV -\varepsilon_{\mathrm{sub}} + E_{\Omega} - E_{\Xi}) \, \big| \bra{\Xi} T_{\mu\nu}(\br) \ket{\Omega} \big|^2,
\end{align}
where $\theta$ is the Heaviside step function, while $D_{\mathrm{tip}}$ and $D_{\mathrm{sub}}$ are the densities of states for the tip and the substrate (which are both assumed to be constant). Next, the differential tunneling conductance takes the form
\begin{align}
G = \frac{\mathrm{d}I} {\mathrm{d}V} &= \frac{2\pi e^2}{\hbar} D_{\mathrm{tip}} D_{\mathrm{sub}} \sum_{\mu\nu} \sum_{\Xi} \int d\varepsilon_{\mathrm{sub}} \, \theta (\varepsilon_{\mathrm{sub}}) \, \delta (eV -\varepsilon_{\mathrm{sub}} + E_{\Omega} - E_{\Xi}) \, \big| \bra{\Xi} T_{\mu\nu}(\br) \ket{\Omega} \big|^2 \nonumber \\
&= \frac{2\pi e^2}{\hbar} D_{\mathrm{tip}} D_{\mathrm{sub}} \sum_{\mu\nu} \sum_{\Xi} \theta (eV + E_{\Omega} - E_{\Xi}) \, \big| \bra{\Xi} T_{\mu\nu}(\br) \ket{\Omega} \big|^2 \\
&= \frac{2\pi e^2}{\hbar} D_{\mathrm{tip}} D_{\mathrm{sub}} \sum_{\mu\nu} \left[ \big| \bra{\Omega} T_{\mu\nu}(\br) \ket{\Omega} \big|^2 + \sum_{\Phi} \theta (eV + E_{\Omega} - E_{\Phi}) \, \big| \bra{\Phi} T_{\mu\nu}(\br) \ket{\Omega} \big|^2 \right], \nonumber
\end{align}
where we separate the general eigenstates $\ket{\Xi}$ into the ground state $\ket{\Omega}$ and the excited states $\ket{\Phi}$ in the last step. Exploiting the identity $\theta (eV + E_{\Omega} - E_{\Phi}) = \int d\omega \, \theta (eV - \omega) \, \delta (\omega + E_{\Omega} - E_{\Phi})$ and noting $E_{\Phi} > E_{\Omega}$, we can then rewrite $G$ into the form of Eq.~(5) in the main text with the elastic and inelastic components given by
\begin{align}
K_{\mathrm{elastic}} = \sum_{\mu\nu} \big| \bra{\Omega} T_{\mu\nu}(\br) \ket{\Omega} \big|^2, \qquad K_{\mathrm{inelastic}} (\omega) = \sum_{\Phi} \delta (\omega + E_{\Omega} - E_{\Phi}) \sum_{\mu\nu} \big| \bra{\Phi} T_{\mu\nu}(\br) \ket{\Omega} \big|^2.
\end{align}
Finally, using Eqs.~(3) and (4) in the main text with $\eta_2 = 0$ and recognizing $\langle \Phi | \Omega \rangle = 0$, the inelastic component becomes
\begin{align}
K_{\mathrm{inelastic}} (\omega) &= \sum_{\Phi} \delta (\omega + E_{\Omega} - E_{\Phi}) \sum_{jk} \sum_{\alpha\beta} \sum_{\mu\nu} T^{\alpha}_{\nu\mu} (\br - \br_j) \, T^{\beta}_{\mu\nu} (\br - \br_k) \bra{\Omega} \sigma^{\alpha}_j \ket{\Phi} \bra{\Phi} \sigma^{\beta}_k \ket{\Omega} \nonumber \\
&= \eta_1^2 \sum_{jk} \sum_{\alpha\beta} T (\br - \br_j) \, T (\br - \br_k) \sum_{\mu\nu} \tau^{\alpha}_{\nu\mu} \tau^{\beta}_{\mu\nu} \sum_{\Phi} \delta (\omega + E_{\Omega} - E_{\Phi}) \bra{\Omega} \sigma^{\alpha}_j \ket{\Phi} \bra{\Phi} \sigma^{\beta}_k \ket{\Omega} \\
&= 2\eta_1^2 \sum_{jk} \sum_{\alpha} T (\br - \br_j) \, T (\br - \br_k) \sum_{\Phi} \delta (\omega + E_{\Omega} - E_{\Phi}) \bra{\Omega} \sigma^{\alpha}_j \ket{\Phi} \bra{\Phi} \sigma^{\alpha}_k \ket{\Omega}, \nonumber
\end{align}
which is exactly Eq.~(6) in the main text with $S^{\alpha\alpha}_{jk} (\omega)$ written in the Lehmann representation [see Eq.~(9) in the main text].

\section{Dynamical spin correlation function}

The site-diluted Kitaev honeycomb model in Eq.~(8) of the main text is exactly solvable. For each flux sector, it reduces to a tight-binding model of free Majorana fermions (i.e., matter fermions) coupled to static $Z_2$ gauge fields:
\begin{align}
\begin{split}
\mathcal{H}_{\mathrm{matter}} = iJ\sum_{\pair{jk}{\al}}u_{\pair{jk}{\al}}c_j c_k+i\kappa\sum_{\langle jkl \rangle_{\al\bt}}u_{\pair{jk}{\al}}u_{\pair{lk}{\bt}}c_j c_l -ih\sum_{j \in \mathbb{D}_{\al}}\tilde{b}^{\al}_{j}c_j.
\end{split}
\end{align}
By properly assigning sublattice indices to the dangling bond fermions $\tilde{b}^{\al}_j$ and treating them as matter fermions $c_j$, this matter-fermion Hamiltonian can be diagonalized in the same way as in the clean model~\cite{Knolle2015}:
\begin{align}
\begin{split}
\mathcal{H}_{\mathrm{matter}} &= \frac{i}{2}
\bg c_A & c_B \ed \bg F & M \\ -M^T & -D \ed \bg c_A \\ c_B \ed = \frac{1}{2} \bg f^{\dg} & f \ed \bg \tilde{h} & \Delta \\ \Delta^{\dg} & -\tilde{h}^{T} \ed \bg f \\ f^{\dg} \ed = \sum_{n} \epsilon_n \left( a^{\dg}_n a_n - \frac{1}{2} \right),
\end{split}
\end{align}
where we apply a complex-fermion transformation and a Bogoliubov transformation:
\begin{align}
\bg f \\ f^{\dg} \ed = \frac{1}{2}\bg \mathbb{1} & i\mathbb{1} \\ \mathbb{1} & -i\mathbb{1}\ed \bg c_A \\ c_B \ed, \quad \bg a \\ a^{\dg} \ed = \bg X^* & Y^* \\ Y & X \ed \bg f \\ f^{\dg} \ed.
\end{align}
In order to calculate the inelastic differential conductance as derived in the previous section, one must then obtain the connected dynamical spin correlation function in terms of fermionic ground-state expectation values:
\begin{align}
S^{\alpha\beta}_{jk} (\omega) = \frac{1} {2\pi} \int_{-\infty}^{+\infty} dt \, e^{i \omega t} \left[ \langle \sg^{\alpha}_j (t) \sg^{\beta}_k (0) \rangle - \langle \sg^{\alpha}_j \rangle \langle \sg^{\beta}_k \rangle \right]. \label{eq:correlator}
\end{align}
As discussed in the main text, these spin correlation functions come in two types: ``bulk'' spin correlation functions that involve non-dangling spin components $\sg^{\alpha}_j$ and $\sg^{\beta}_k$, and ``dangling'' spin correlation functions that involve dangling spin components $\tilde{\sg}^{\alpha}_j$ and $\tilde{\sg}^{\beta}_k$. (Note that the spin correlation function vanishes if one spin component is dangling while the other one is not.)

\subsection{Bulk spin correlation functions}
Each bulk spin correlation function can be mapped to a quantum quench problem where the perturbation involves a bond-flipping process. The ground state is a product state of the ground-state flux sector and the corresponding matter-fermion vacuum: $\ket{\Omega} = \ket{F}\otimes \ket{M}$. Note that, in the presence of vacancies, $\ket{F}$ is not required to be the zero-flux sector~\cite{Willans2010, Willans2011, Kao2021vacancy}. The first term of the spin correlation function in Eq.~(\ref{eq:correlator}) is given by~\cite{Knolle2015}
\begin{align}
\expval{\sg^{\al}_j(t)\sg^{\bt}_k(0)} = \xi_{jk} \dt_{\al\bt} \dt_{\langle jk\rangle_{\al}} \bra{M}e^{i\mathcal{H}_{\mathrm{matter}}t}c_j e^{-i\mathcal{H}_{\mathrm{matter}}^{\prime}t}c_k\ket{M},
\end{align}
where $\mathcal{H}_{\mathrm{matter}}^{\prime}$ is a modified matter-fermion Hamiltonian in which the gauge field $u_{\pair{jk}{\al}}$ is flipped with respect to the ground state, while $\dt_{\langle jk\rangle_{\al}}$ reflects that the spin correlation function is ultra-short-ranged due to the corresponding change in the flux sector~\cite{Baskaran2007, Knolle2014}. We also introduce a shorthand notation for different combinations of the sublattices,
\begin{align}
    \xi_{jk} = \bg 1 & -i \\ i & 1 \ed,
\end{align}
where the first row/column refers to the A sublattice and the second row/column refers to the B sublattice. Taking the adiabatic approximation that is known to be quantitatively correct~\cite{Knolle2015}, and noting that $\langle \sg^{\al}_j\rangle = 0$ for each non-dangling spin component $\sg^{\al}_j$, the spin correlation function in Eq.~(\ref{eq:correlator}) can then be written in the Lehmann representation as
\begin{align}
S^{\al\bt}_{jk}(\omega) = \xi_{jk} \dt_{\al\bt} \dt_{\langle jk\rangle_{\al}} \sum_{\ld}\bra{M'}c_j (a_{\ld}')^{\dg}\ket{M'}\bra{M'}a_{\ld}'c_{k}\ket{M'}\dt\left[ \omega - (E_{\Phi_{\ld}} - E_{\Omega})\right],
\end{align}
where $\ket{M'}$ is the vacuum of the matter fermions $(a_{\ld}')^{\dg}$ that correspond to the modified flux sector $\ket{F'}$ containing a pair of flux excitations around the bond $\pair{jk}{\al}$, while $E_{\Omega}$ and $E_{\Phi_{\ld}}$ are the energies of the ground state $\ket{\Omega} = \ket{F} \otimes \ket{M}$ and the excited state $\ket{\Phi_{\ld}} = \ket{F'} \otimes (a_{\ld}')^{\dg} \ket{M'}$, respectively. The two-fermion expectation values in this expression are straightforward to evaluate.

\subsection{Dangling spin correlation functions}
Since a dangling spin operator can be decomposed into a matter fermion and a dangling bond fermion, $\tilde{\sg}^{\al}_j = i\tilde{b}^{\al}_j c_j$, the flux sector does not change when $\tilde{\sg}^{\al}_j$ is acting on $\ket{\Omega}$. Thus, for a dangling spin correlation function, the Lehmann representation of the first term in Eq.~(\ref{eq:correlator}) is simply given by
\begin{align}
\expval{\tilde{\sg}^{\al}_j(t)\tilde{\sg}^{\bt}_k (0)} = -\sum_{\Phi}e^{i(E_{\Omega}-E_{\Phi})t} \bra{\Omega}\tilde{b}^{\al}_j c_j\ket{\Phi}\bra{\Phi}\tilde{b}^{\bt}_k c_k\ket{\Omega}.
\end{align}
The excited state $\ket{\Phi}$ with energy $E_{\Phi}$ can be either a zero-fermion state or a two-fermion state. For the zero-fermion contribution, $\ket{\Phi} = \ket{\Omega}$ and $E_{\Phi} = E_{\Omega}$. Therefore,
\begin{align}
\left.\expval{\tilde{\sg}^{\al}_j(t)\tilde{\sg}^{\bt}_k (0)}\right|_{\textrm{0-fermion}} = -\bra{\Omega}\tilde{b}^{\al}_j c_j\ket{\Omega}\bra{\Omega}\tilde{b}^{\bt}_k c_k\ket{\Omega} = \langle\tilde{\sg}^{\al}_j\rangle\langle\tilde{\sg}^{\bt}_{k}\rangle,
\end{align}
which simply cancels the second term of Eq.~(\ref{eq:correlator}). For the two-fermion contribution, $\ket{\Phi} = a^{\dg}_{\gm}a^{\dg}_{\dt}\ket{\Omega}$ and $E_{\Phi} = E_{\Omega} + \epsilon_{\gamma} + \epsilon_{\delta}$, hence we immediately obtain
\begin{align}
\left.\expval{\tilde{\sg}^{\al}_j(t)\tilde{\sg}^{\bt}_k (0)}\right|_{\textrm{2-fermion}} &= -\sum_{\gm\dt}e^{-i(\epsilon_{\gm}+\epsilon_{\dt})t}\bra{\Omega}\tilde{b}^{\al}_j c_j a^{\dg}_{\gm}a^{\dg}_{\dt}\ket{\Omega}\bra{\Omega}a_{\dt}a_{\gm}\tilde{b}^{\bt}_{k}c_k\ket{\Omega} \\
&= -\sum_{\gm\dt}e^{-i(\epsilon_{\gm}+\epsilon_{\dt})t}\bra{M}\tilde{b}^{\al}_j c_j a^{\dg}_{\gm}a^{\dg}_{\dt}\ket{M}\bra{M}a_{\dt}a_{\gm}\tilde{b}^{\bt}_{k}c_k\ket{M}. \nonumber
\end{align}
Therefore, the spin correlation function in Eq.~(\ref{eq:correlator}) can be calculated in terms of four-fermion expectation values:
\begin{align}
S^{\al\bt}_{jk}(\omega) = -\sum_{\gm\dt}\bra{M}\tilde{b}^{\al}_j c_j a^{\dg}_{\gm}a^{\dg}_{\dt}\ket{M}\bra{M}a_{\dt}a_{\gm}\tilde{b}^{\bt}_{k}c_k\ket{M}\dt\left[ \omega-(\epsilon_{\gm}+\epsilon_{\dt}) \right].
\end{align}
We remark that, for the dangling spin correlation functions, there is no selection rule from flux excitations, which means that the non-local dangling spin correlations are potentially observable in a multi-tip STM setup \cite{takahashi2023nonlocal}. However, due to the localized nature of the low-energy modes, these non-local correlations, or \textit{inter-vacancy correlations}, are much smaller than the \textit{intra-vacancy correlations} considered in this work.

\section{Density of states and inverse participation ratio}

\begin{figure}
\includegraphics[width=1.0\columnwidth]{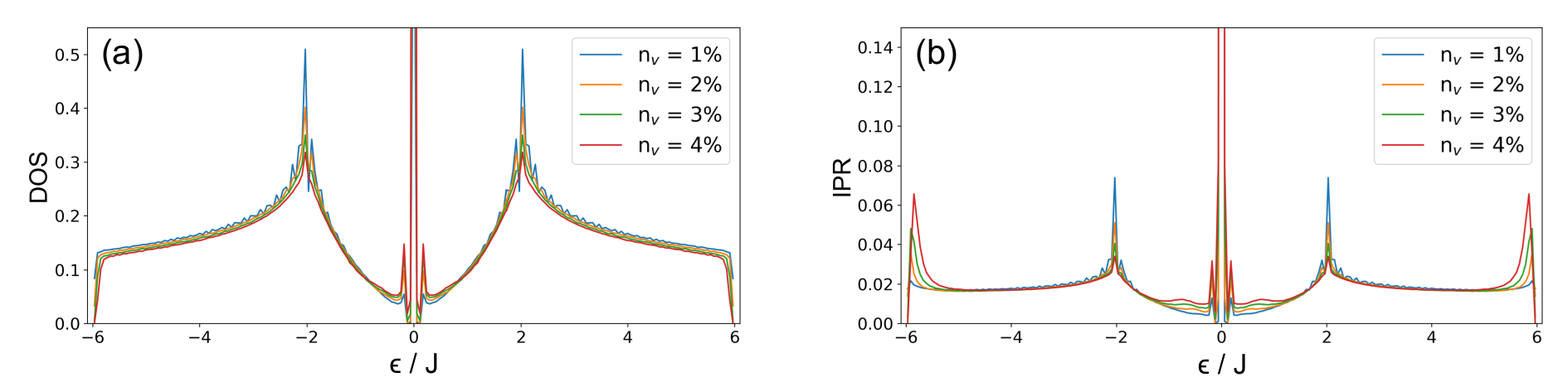}
     \caption{\label{fig:FigureS2} Disorder-averaged results for (a) the density of states (DOS) and (b) the inverse participation ratio (IPR) for $\mathcal{H}_{\mathrm{Kitaev}}$ in Eq.~(8) of the main text with $\kp = 0.02J$ and $h = 0.05J$. The positive-energy spectrum is mirrored to the negative-energy side in order to highlight the peaks close to zero energy. Each curve is averaged over 500 disorder realizations in a system containing $40\times 40$ unit cells. The DOS for each realization is calculated as a histogram of energy eigenvalues $\epsilon_n$. For each eigenmode $n$ with wavefunction amplitude $\phi_{n,j}$ on site $j$, the IPR is calculated as $\sum_j |\phi_{n,j}|^4$. Higher magnitude of IPR indicates higher level of localization for the given eigenmode.}
\end{figure}

In the presence of a finite density of quasivacancies or true vacancies, the low-energy density of states (DOS) of the Kitaev honeycomb model changes drastically from linear behavior in the clean model into a set of distinctive peaks in the site-diluted model \cite{Kao2021vacancy,Kao2021localization}. This accumulation of states comes from the vacancy-induced localized modes. The central sharp peak in Fig.~\ref{fig:FigureS2}(a) represents the eigenmodes with mostly $\tilde{b}^{\al}_{j}$ character. On the other hand, the satellite peak away from zero energy can be attributed to localized modes with mostly $c_j$ character. The low-energy response shown in the main text can then be understood by noting that each dangling spin component creates two fermion excitations. Further, by calculating the inverse participation ratio (IPR), which is an indicator of wavefunction spread in real space for each eigenmode, we can show that the dangling $\tilde{b}^{\al}_{j}$ modes are significantly more localized than all the other modes of the system [see Fig.~\ref{fig:FigureS2}(b)]. Therefore, these low-energy localized modes are hardly detected in transport properties but can be unveiled by local probes such as STM.

%\bibliographystyle{apsrev4-1}
%\bibliography{snakeref}
\bibliography{STM_refs.bib}